# Climate land use and other drivers' impacts on island ecosystem services: a global review

*Aristides Moustakas[1,*], Shiri Zemah-Shamir[2], Mirela Tase[3], Savvas Zotos[4,] Nazli Demirel[5], Christos Zoumides[6], Irene Christoforidi[7], Turgay Dindaroglu[8], Tamer Albayrak[9,10], Cigdem Kaptan Ayhan[11], Mauro Fois[12], Paraskevi Manolaki[4], Attila D. Sandor[13,14], Ina Sieber[15], Valentini Stamatiadou[16], Elli Tzirkalli[4], Ioannis N. Vogiatzakis[4,17], Ziv Zemah-Shamir[18], George Zittis[19, *]*

[1]Natural History Museum of Crete, University of Crete, Heraklion, Greece
[2]School of Sustainability, Interdisciplinary Center (IDC), Reichman University, Herzliya, Israel
[3]Department of Tourism, Aleksander Moisiu University, Durrës, Albania
[4]Faculty of Pure and Applied Sciences, Open University of Cyprus, Nicosia, Cyprus
[5]Institute of Marine Sciences and Management, Istanbul University, Turkey
[6]Energy, Environment and Water Research Center (EEWRC), The Cyprus Institute, Nicosia, Cyprus
[7]Department of Plant Production Experimental farm, Hellenic Mediterranean University, Heraklion, Greece
[8]Department of Forest Engineering, Karadeniz Technical University, Trabzon, Turkey
[9]Department of Biology, Budur Mehmet Akif Ersoy University, Burdur, Turkey
[10]Department of Mathematics and Science Education, Buca Faculty of Education, Dokuz Eylül University, İzmir, Turkey
[11]Department of Landscape Architecture, Canakkale Onsekiz Mart University, Canakkale, Turkey
[12]Department of Life and Environmental Sciences, University of Cagliari, Cagliari, Italy
[13]HUN-REN-UVMB Climate Change: New Blood-Sucking Parasites and Vector-Borne Pathogens Research Group, Budapest, Hungary
[14]University of Agricultural Sciences and Veterinary Medicine Cluj-Napoca, Cluj-Napoca, Romania
[15]Kassel Institute for Sustainability, Kassel University, Kassel, Germany
[16] Department of Marine Sciences, University of the Aegean, Mytilene, Greece
[17] Department of Soil, Plant and Food Sciences, University of Bari Aldo Moro, Bari, Italy
[18]Department of Marine Biology, Leon H. Charney School of Marine Sciences, University of Haifa, Haifa, Israel
[19]Climate and Atmosphere Research Center (CARE-C), The Cyprus Institute, Nicosia, Cyprus

*Corresponding Authors:
Aristides (Aris) Moustakas: arismoustakas@gmail.com
George Zittis: g.zittis@cyi.ac.cy

**Abstract**

Islands are diversity hotspots and vulnerable to environmental degradation, climate variations, land use changes and societal crises. These factors can exhibit interactive impacts on ecosystem services. The study reviewed a large number of papers on the climate change-islands-ecosystem services topic worldwide. Potential inclusion of land use changes and other drivers of impacts on ecosystem services were sequentially also recorded. The study sought to investigate the impacts of climate change, land use change, and other non-climatic driver changes on island ecosystem services. Explanatory variables examined were divided into two categories: environmental variables and methodological ones. Environmental variables include sea zone geographic location, ecosystem, ecosystem services, climate, land use, other driver variables, Methodological variables include consideration of policy interventions, uncertainty assessment, cumulative effects of climate change, synergistic effects of climate change with land use change and other anthropogenic and environmental drivers, and the diversity of variables used in the analysis. Machine learning and statistical methods were used to analyze their effects on island ecosystem services. Negative climate change impacts on ecosystem services are better quantified by land use change or other non-climatic driver variables than by climate variables. The synergy of land use together with climate changes is modulating the impact outcome and critical for a better impact assessment. Analyzed together, there is little evidence of more pronounced for a specific sea zone, ecosystem, or ecosystem service. Climate change impacts may be underestimated due to the use of a single climate variable deployed in most studies. Policy interventions exhibit low classification accuracy in quantifying impacts indicating insufficient efficacy or integration in the studies.

## 1. Introduction

The accelerating pace of global change, marked by profound climatic and land-use changes, poses unprecedented challenges to biodiversity conservation and sustainable development (IPBES, 2019; WWF, 2022). Among the most vulnerable are the world's islands, which, despite covering a modest fraction of the Earth's surface, harbor a disproportionate share of its biodiversity. Over one-third of the global biodiversity hotspots for conservation priority consist mainly or entirely of islands, while almost all tropical islands fall into such a hotspot region (Myers et al., 2000). However, there are knowledge gaps regarding island ecosystem services and their response to climatic and land use changes (Sieber et al., 2018), and thus, science-policy assessments are hindered by such a lack of information (Cámara-Leret and Dennehy, 2019).

Ecosystem services are the contributions that ecosystems make to human well-being (MEA, 2005; Edens et al., 2022; UN et al., 2024). Under the Common International Classification of Ecosystem Services (CICES v5.1) framework (Haines-Young & Potschin, 2012), these are structured into provisioning, regulation and maintenance, and cultural services. They are supported by essential ecosystem processes and functions, such as photosynthesis and nutrient cycling (Hassan et al., 2005; UN et al., 2024); these processes are indispensable for sustaining ecosystems, in CICES they are classified as supporting functions rather than final services. Reciprocally, ecosystem services delivery strongly depends upon climate conditions and land use practices, and can be negatively affected by anthropogenic disturbances, such as deforestation, pollution, and urban expansion.

Islands rely heavily on ecosystem services derived from their terrestrial, coastal and marine environments. In addition, these island ecosystems may also deliver services beyond their immediate terrestrial boundaries. For example, marine and coastal ecosystems provide final ecosystem services such as nursery habitats (termed 'lifecycle maintenance' in CICES v5.1), carbon sequestration and experiential recreation opportunities, which benefit both local biodiversity and distant societies (e.g., visiting tourists). In temperate regions, ocean ecosystems can be more sensitive to climatic change than terrestrial ones (Antão et al., 2020). However, it is not well known whether this applies when comparing marine and terrestrial ecosystems of islands across regions, leaving a critical gap in science-policy dialogues; in the EU ecosystem assessment islands are simply mentioned in the annex (Maes et al., 2021). Are there any particular ecosystem services more impacted than others?

Climate has a direct impact on ecosystems and the services they provide (IPBES, 2019; WWF, 2022, 2020). Island ecosystems are particularly susceptible to the effects of climate change, including rising temperatures, sea levels, and extreme events (Garlati, 2013; IPCC, 2023). Key parameters most primarily affecting ecosystems are temperature (e.g. air, land, and ocean), components of the hydrological cycle (e.g. precipitation and soil moisture), and oceanic properties such as acidity. Changes in the mean state, variability, and seasonality of these parameters can significantly impact ecosystems (Gruber et al., 2021; Ruthrof et al., 2021) and the services they provide. Is there a climate variable with more pronounced impacts on ecosystem services?

Various other drivers affect ecosystem services, in synergy with changes in climate. The integrated study of changes with land use is crucial. Transitions from natural land cover to artificial is related with increased local temperatures (Zhao et al., 2020). Land use changes also influence climate change by moving from low-carbon activities to more impactful ones (Pielke Sr et al., 2011) Pielke Sr et al., 2011). The impacts of land use changes on ecosystem services are non-ubiquitous (Médail, 2017). For instance, while transitions to agricultural land may enhance food-provisioning services, they can detrimentally affect climate-regulation services (Azadi et al., 2021; Viana et al., 2022). Conversely, certain strategies, such as habitat protection, rewilding, and reforestation, can have positive impacts on biodiversity, and the provision of ecosystem services (Andres et al., 2023; Xu et al., 2017). Findings quantifying threats of climate, land use, and other drivers rank land use changes as the biggest threat to biodiversity across continents (IPBES, 2019; WWF, 2020), while this is not meant to downplay the impacts of climate. These conclusions are predominantly based on terrestrial data, often lacking the fine-scale resolution necessary to fully understand island-specific dynamics.

Box 1| **Island ecosystem services, climate change, land use changes, and other drivers**

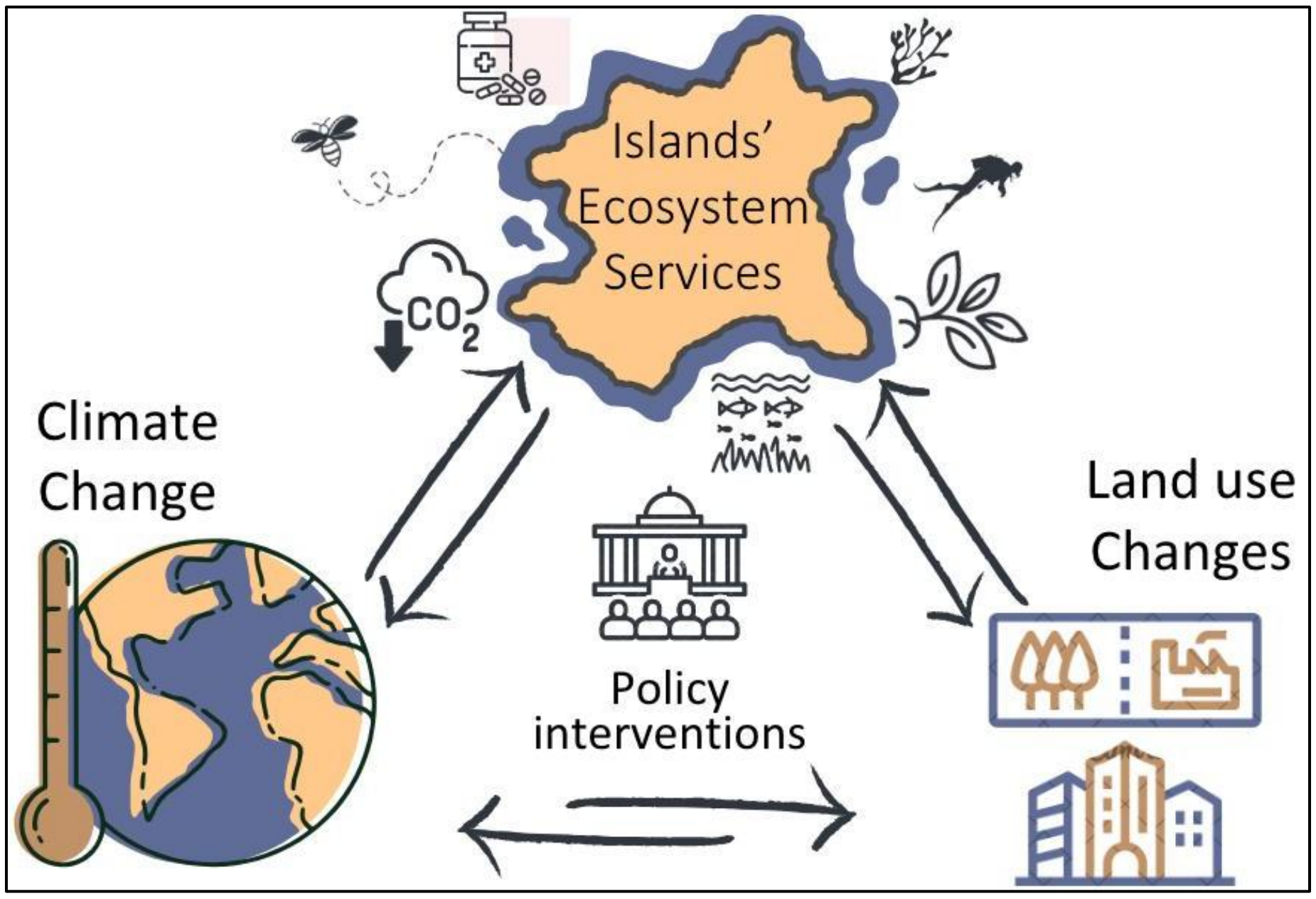

Ecosystem services have always had a strong interplay between climate and land use changes (Behringer, 2010). One of the worst human population collapses occurred during the early 14th century in northern Europe; the 'Great Famine' was the consequence of the dramatic effects of climate deterioration (Lima, 2014). In a warming climate phase, the Vikings established a colony in Greenland in the late 10th century during a local climatic peak and left thereafter. A decline in the capacity of coral reefs to provide ecosystem services is recorded due to climate change (Eddy et al., 2021).

However, not all environmental changes can be simply attributed to climate. The Marshall Islands, one of the lowest-lying island nation states in the world, is vulnerable to sea level rise, flooding and the associated impacts on soil and water salinity (van der Geest et al., 2020). However, the majority of the islands' population mention education, health care, work and family visits as migration drivers, and only a few mention climate impacts or environmental change (van der Geest et al., 2020). Similarly, in Papua New Guinea, climate change was a less critical driver of change than human population growth (Butler et al., 2014). Ultimately the level of complexity and diversity of assessment features in terms of ecosystem services examined, climate variables, ecosystem type, time of the study, spatial scale of assessment, geographic location, inclusion of other non-climatic drivers and the number of features of each driver, are impacting the output and are associated with uncertainty (Walther et al., 2025). Therefore, there is a need for a holistic multi-factor uncertainty assessment.

Various other drivers affect ecosystem services, aside from changes in climate or land use. These include demographic growth, economic or technological advancements that may intensify natural resources use and land degradation, including soil contamination, pollution, and erosion. Furthermore, there are adverse effects of invasive alien species (Bjarnason et al., 2017), diseases and pests (Boyd et al., 2013), partly through interactions with land use and climatic changes. Without adaptive management interventions, these pressures can trigger permanent ecological shifts with very negative or unchartered impacts on ecosystem services (Harris et al., 2018). To mitigate anthropogenic disturbances against biodiversity, policy and management interventions are introduced at global, regional or local levels. In the European Union (EU), for instance, the new biodiversity strategy aims to increase the protected areas to at least 30 % of EU land by 2030 (European Commission, 2021). Rewilding the EU is also a core part of the strategy (European Commission, 2021). On top of this plan, a second pioneering proposal aims to restore 80% of European habitats that are in poor condition (European Commission, 2022). The effectiveness of such bold interventions and, therefore, whether or how targets should be adapted to islands, is not well known. Does the island area and geographic location matter? Climatic changes are not always linear or uniform in direction and magnitude. For example, recent warming in the Mediterranean (Urdiales-Flores et al., 2023), and the polar regions (Clem et al., 2020) is reportedly faster than the global average and other parts of the planet. Are there islands in particular ocean zones with higher impacts on ecosystem services?

Assessments often focus on a specific ecosystem service because it is critical to islands' functioning or human well-being. However, a broader or simultaneous investigation may reveal divergent impacts of across ecosystem services (He and Silliman, 2019). Additionally, considering multiple ecosystem types (marine, terrestrial, freshwater) may uncover diverse impacts. Likewise, studying a single climatic feature may underestimate the impacts on ecosystem services compared to assessments that include multiple parameters. The inclusion of non-climatic drivers may act synergistically or antagonistically on island ecosystem services (Redlich et al., 2022). Similarly to the climate features examined, the diversity of other drivers may influence the type of effect (Homet et al., 2024). When one makes a single hypothesis, it is more likely that it becomes attached to it (Platt, 1964). Admittedly, the issue of climate, land use, and other driver changes on island ecosystem services is a complex one (Box 1). Is this complexity embraced in the analysis? Is the number of features examined determining the output of impacts on ecosystem services? To what extent is uncertainty incorporated in the analysis?

Here, we address these gaps by conducting a global review and extracting a comprehensive set of features related to climate change, land use change, and other drivers impacting island ecosystem services. Utilizing advanced machine learning techniques, we simultaneously quantify the effects of these features across multiple ecosystem services, offering a better understanding of their interactions and relative impacts. Our analysis embraces the complexity and diversity of island ecosystems around the world, incorporating multiple parameters. This study represents the first global-scale, quantitative synthesis of its kind, providing robust and up-to-date evidence on the interplay of various drivers on island ecosystem services.

The questions addressed in terms to environmental and physical assessment include: (1) Are climate change impacts on ecosystem services more pronounced on a specific ecosystem type or ecosystem service? (2) Is there a climate variable, a land use type, or other driver variable with more pronounced impacts on ecosystem services? (3) Is there a geographic location in terms of sea zone with more pronounced impacts of climate change on ecosystem services? The questions addressed regarding the potential impact recorded based on the methodological approach deployed include: (4) To what extend is consideration of synergistic climate change effects as well as interactions with land use changes and other drivers influencing the impact outcome? (5) To what extend is the complexity and diversity of analysis in terms of number of variables and features included affecting the impact outcome? (6) To what extend is consideration of uncertainty, policy or management interventions affecting the impact of climate change on ecosystem services outcome?

## 2. Methods

### 2.1. Literature review and screening

#### *2.1.1. Islands and sea zones*

Sea zones were defined using the Major Fishing Areas as classified by the Food and Agriculture Organisation (FAO) of the United Nations listing 19 major sea zones (United Nations 1982; Table 1). These are arbitrary areas, the boundaries of which were determined in consultation with international fishery agencies on various considerations, including (i) the boundary of natural regions and the natural divisions of oceans and seas, (ii) the boundaries of adjacent statistical fisheries bodies already established in inter-governmental conventions and treaties (iii) existing national practices, (iv) national boundaries, (v) the longitude and latitude grid system. (vi) the distribution of the aquatic fauna, and (vii) the environmental conditions within an area. For this review, we consider islands to be areas of land surrounded by these waters, thus excluding inland islands (i.e. islands in rivers, lakes or any other inland island formation). In particular, we follow the international standard for the definition of what an island is, according to the United Nations Convention on the Law of the Sea. The key features that characterize islands and distinguish them from "rocks" are that an island is a naturally formed area of land, surrounded by water, which is above water at high tide. In contrast with "rocks", islands can sustain human habitation or economic life of their own and have exclusive economic zones or continental shelves (United Nations 1982). We added an extra sea zone coded as "Global" in case a global assessment was performed (Table 1). Classification is not mutually exclusive and papers studying several sea zones were classified in each of them.

2.1.2. *Literature screening*

The initial literature search used the following terms and boolean search operators: *("ecosystem service*" OR "ecosystem good*") AND (climate* NEAR chang*) AND (island* OR islet* OR archipelag*)*. These were applied on a topic search in Scopus and ISI Web of Science Core Collection. The final search was conducted in September 2024 and includes papers published till the end of December 2023. The search was restricted to include only peer-reviewed scientific articles, including original research items, letters and review studies. We did not include more specific ecosystem services search terms such as 'crops' or 'fisheries' that could bias the results toward these services and return an impractical number of papers. The same applies to specific climate change phenomena e.g., 'sea-level rise' or 'global warming'.

Literature screening followed the PRISMA method (Page et al. 2021). Applying the search protocol regarding climatic change and ecosystem services and island, resulted in n=587 publications from Web of Science and n=591 from Scopus. In this step, we also applied a first screening for removing n=234 duplicates. Non-peer-reviewed n=110 studies, were also excluded as well as n=4 studies that could not be retrieved (inaccessible). Irrelevant studies that had no island focus n=273, or no relevance to climate change and/or ecosystem services n=331 were also excluded. No island focus, refer to papers not studying islands or studying inland islands. For example, "rocks" were excluded based on the definition above; similarly, inland islands of lakes or rivers were also excluded. No climate change focus refers to numerous cases where the term "island" appeared but was referring at a different concept, such as "Urban Heat Island", a phenomenon that refers to urban areas being significantly warmer than their surrounding rural areas due to human activities. The term "island" also appeared in studies regarding regions with toponyms including the string "Island", that, however, were not islands according to the definition we followed. These cases were all excluded. This screening resulted in a total of 226 publications, including 25 review studies. An overview of the literature screening process is provided in Fig. 1. The inclusion of review articles can potentially risk double-counting some research studies. While some review articles may reference others within the collection, many provide essential novel insights or re-analyses that we felt were important to include. Given the geographic diversity, the interdisciplinary aspects, and the limited number of review articles compared to the overall sample, we are confident that their inclusion enhances our findings without significantly altering our conclusions.

It is important to note that "land use changes" or "other drivers" are not explicitly incorporated in the publication search methodology and that the number of papers would have been considerably lower had it been included. Our literature search did not include the term "land use change" as we wanted to avoid excluding several relevant studies. In addition, "other non-climatic drivers" include a long list of potential topics such as invasive species, human population growth, economic growth or austerity among others that are not straightforward in search terms. Instead, during the screening phase and within the recording parameters template (Table 1), land use change or other drivers' inclusion and if included the features or variables investigated were recorded.

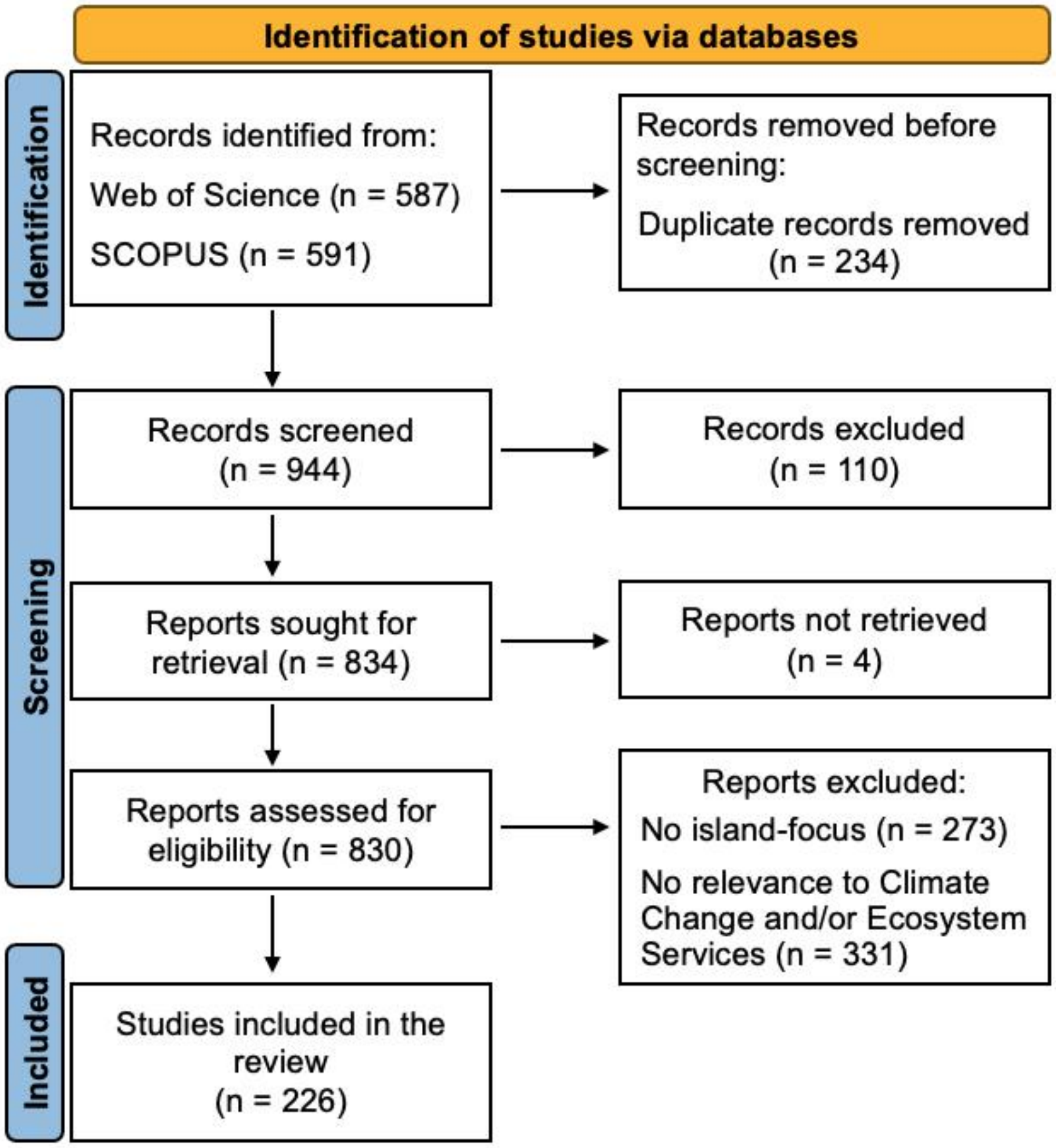

**Figure 1.** An overview of the literature screening process following the PRISMA method.

*2.2. Recording impact*

For each publication, we recorded the impact of climate change as well as land use change and other driver were included on ecosystem service(s). The impact was quantified by the reader's evaluation as negative, unclear, neutral, or positive with mutually exclusive clustering. Impact evaluation is derived by the outcome of the driver (climate change, and if included land use change, or other driver), on ecosystem services based on the words or terms that the authors used in the publication and it is quantified by the expert's subjective recording. Each publication was read by at least one contributing author, while a random 20% of publications were read by two co-authors for cross-validation. Differences in recording variables were negligible. Negative impacts were recorded in cases where climate change or land use change or other driver was negatively associated with ecosystem services on the study island. Unclear impacts were recorded in studies where the relationship between climate or land use change was not clearly concluded or stated, neutral when climate or land use had no effect on ecosystem services while positive when the change enhanced ecosystem services.

*2.3. Environmental variables*

The explanatory independent variables included sea zone, ecosystem type, ecosystem service type, climate variable, land use variable, and other driver variables listed in Nr. 1 – 57 in Table 1. To ensure clarity and consistency in defining ecosystem services relevant to island studies, we adapted a set of 12 service classes informed by CICES v5.1 (https://cices.eu/). Table 1 below shows the resulting classification. Each variable consisted of a binary (0, 1) record. Doing so, classifications of papers can include several categories simultaneously. For example, if ecosystem services studied were bioremediation (e.g., in wetlands) that falls in the category 'waste and toxin mediation' AND plant genes for biotech (e.g., synthetic drugs) under the 'genetic and biochemical resources', these reciprocal columns will be scored with '1' while all other columns of ecosystem services with '0' for that publication. Keeping a binary response for each variable is known to minimize bias in the analysis (Strobl et al., 2007).

2.4. *Methodological variables*

Methodological variables included the richness of ecosystem types studied ('Ecosystem types', sum of variables 21-23 in Table 1), richness of ecosystem services studied ('ES richness', sum of variables 24-35, richness of climate features studied ('CC richness', sum of variables 36-40), richness of other driver features ('OD richness', sum of variables 41-49), and the richness of land use features studied ('LU richness', sum of variables 50-57). In addition, in each paper it was recorded as a binary variable whether the impact of the other driver was cumulatively or individually assessed ('ODIA individual', 'ODIA cumulatively'), if the study included synergistic or individual assessment between climate change and the other driver's impacts ('CC-OD synergistic', 'CC-OD individual'), if uncertainty assessment was performed ('Uncertainty') and if policy interventions were considered ('Interventions').

*2.5. Machine Learning*

In order to statistically explain the impacts of climate change drivers on island ecosystem services and the explanatory variables listed in Table 1, Random Forests (RF) classifiers were used (Breiman 2001). RFs employ boosting (Schapire et al., 1998) and bagging algorithms (bootstrap aggregating (Breiman 1996)) of the Classification and Regression Tree (CART (Breiman 2017)) model and use a more random but more efficient node splitting strategy than standard CARTs (Liaw and Wiener 2002). RFs have been employed in a wide variety of classification and prediction problems as they rank well among the most efficient analytic tools for extracting information in noisy, complex, potentially correlated, and high dimensional datasets (Scornet et al., 2015; Wager et al., 2014). Analysis was conducted in 'skikit-learn' library in Python.

*2.6. Environmental analysis*

The dependent variable was the impact of climate change on island ecosystem services with feasible values 'negative' n=118, 'unknown' n= 66, 'neutral' n=37, or 'positive' n=6. Each outcome was analyzed separately (four sets of analysis one for each outcome, with the same independent variables and corresponding scores), to quantify the specific combinations of variables resulting in such an output. We trained and tested an RF classifier, using 10 different random states. The data split partitioning explored included a 70-30% training to test data partitions. Sensitivity analysis was performed by changing the training to testing data partition by up to ±10%. . We did not use several machine learning methods simultaneously and compare their accuracy because the aim of this work is to explain rather than predict in novel conditions and compare optimal predictive tools and their accuracy. The accuracy is defined as the number of correct bin classifications divided by the total number of instances in the dataset.

**Table 1.** List of environmental variables used in the analysis. All variables were recorded each in a unique column with binary scores of 0 (no) or 1 (yes).

| Var Nr | Sea zone | Value |
|---|---|---|
| 1 | | ARCTIC |
| 2 | | ATLANTIC NORTHWEST |
| 3 | | ATLANTIC NORTHEAST |
| 4 | | ATLANTIC WESTERN CENTRAL |
| 5 | | ATLANTIC EASTERN CENTRAL |
| 6 | | MEDITERRANEAN |
| 7 | | ATLANTIC SOUTHWEST |
| 8 | | ATLANTIC SOUTHEAST |
| 9 | | INDIAN WESTERN |
| 10 | | INDIAN EASTERN |
| 11 | | PACIFIC NORTHWEST |
| 12 | | PACIFIC NORTHEAST |
| 13 | | PACIFIC WESTERN CENTRAL |
| 14 | | PACIFIC EASTERN CENTRAL |
| 15 | | PACIFIC SOUTHWEST |
| 16 | | PACIFIC SOUTHEAST |
| 17 | | ATLANTIC ANTARCTIC |
| 18 | | INDIAN ANTARCTIC |
| 19 | | PACIFIC ANTARCTIC |
| 20 | | GLOBAL |
| | **Ecosystem Type** | |
| 21 | | Marine |
| 22 | | Freshwater |
| 23 | | Terrestrial |
| | **Ecosystem Service** | |
| 24 | | ES: Food & Nutrition Provision |
| 25 | | ES: Biotic & Abiotic Material Provision |
| 26 | | ES: Freshwater Provision & Regulation |
| 27 | | ES: Genetic & Biochemical Resources |
| 28 | | ES: Climate & Atmospheric Regulation |
| 29 | | ES: Carbon Storage & Sequestration |
| 30 | | ES: Waste & Toxin Mediation |
| 31 | | ES: Soil Formation & Erosion Control |
| 32 | | ES: Biological Control & Pollination |
| 33 | | ES: Physical & Experiential Recreation |
| 34 | | ES: Aesthetic & Inspirational Value |
| 35 | | ES: Cultural & Spiritual Significance |
| | Climate impact driver | |
| 36 | | CC: Temperature |
| 37 | | CC: Precipitation |
| 38 | | CC: Sea level rise |
| 39 | | CC: Ocean acidification |
| 40 | | CC: Extremes |
| | Other driver | |
| 41 | | OD: Land use |
| 42 | | OD: Alien species |
| 43 | | OD: Population growth |
| 44 | | OD: Economic factors |
| 45 | | OD: Pollution |
| 46 | | OD: Diseases and pests |
| 47 | | OD: Land degradation |
| 48 | | OD: Technological advancements |
| 49 | | OD: Management/policy |
| | Land Use | |
| 50 | | LU: Urban expansion |
| 51 | | LU: Mining |
| 52 | | LU: Deforestation |
| 53 | | LU: Reforestation |
| 54 | | LU: Rewilding |
| 55 | | LU: Wetland modification |
| 56 | | LU: Coastal degradation |
| 57 | | LU: Habitat protection |

### *2.7. Methodological analysis*

The dependent variable was the impact of climate change on island ecosystem services with feasible values 'negative', 'unknown', 'neutral', or 'positive'. Independent variables included the list of variables under "methodological variables" section. Outcome was analyzed simultaneously (one set of analysis with the dependent variable exhibiting four scores). This was conducted to quantify the classification accuracy of methodological approaches and variables in detecting climate change impacts, regardless of what the outcome of the impact was. All dependent variables were normalized in [0, 1] interval so that they are on the same scale and avoid potential biases of richness-related variables that have larger scores than 1. We trained and tested an RF classifier, using 10 different random states. The data split partitioning explored included a 70-30% training to test data partitions. Sensitivity analysis was performed by changing the training to testing data partition by up to ±10%. . Analysis was conducted in 'skikit-learn' library in Python.

### 2.8. Variable importance

In order to quantify the importance of each of the explanatory variables we calculated the mean decrease in impurity for each variable (Pedregosa et al., 2011). We used the 'skikit-learn' embedded feature importance method in Python software, that evaluates the impurity importance based feature of the forest, along with their inter-trees variability (Pedregosa et al., 2011); see also available code in (Moustakas and Davlias 2021). The relative variable importance chart was used to measure the relative importance of each factor in the impact assessment outcome. It shows predictors in descending order of their impact on model enhancement from all basis functions (Venkateswarlu and Anmala 2024). The chart standardizes importance values for easier interpretation. Relative importance is defined as the percentage improvement relative to the most important predictor, which has an importance of 100%. (Venkateswarlu and Anmala 2024) It is calculated by dividing each variable's importance score by the largest value and multiplying by 100%.

## 3. Results

## 3.1. Environmental analysis

### *3.1.1. Geographic location*

Climate change impacts were not significantly different per sea zone except the Arctic where the effect was marginal more positive (Fig. 2a). Climate change impacts per sea zone exhibit increased variance and confidence intervals in Atlantic South East, Pacific Eastern Central, while smaller variances are recorded in Atlantic Western Central and the Mediterranean (Fig. 2a).

### *3.1.2. Ecosystem type*

Negative impacts are predominantly reported in all three ecosystems, marine, terrestrial, and freshwater (Fig. 2b). Frequencies of negative impacts are relatively higher in marine but in general similar across ecosystems (Fig. 2b). Unclear impacts are also similar across ecosystems and peaking at terrestrial ecosystems (Fig. 2b). Neutral impacts are most frequently reported at freshwater ecosystems (Fig. 2b). Very few positive impacts are reported (Fig. 2b).

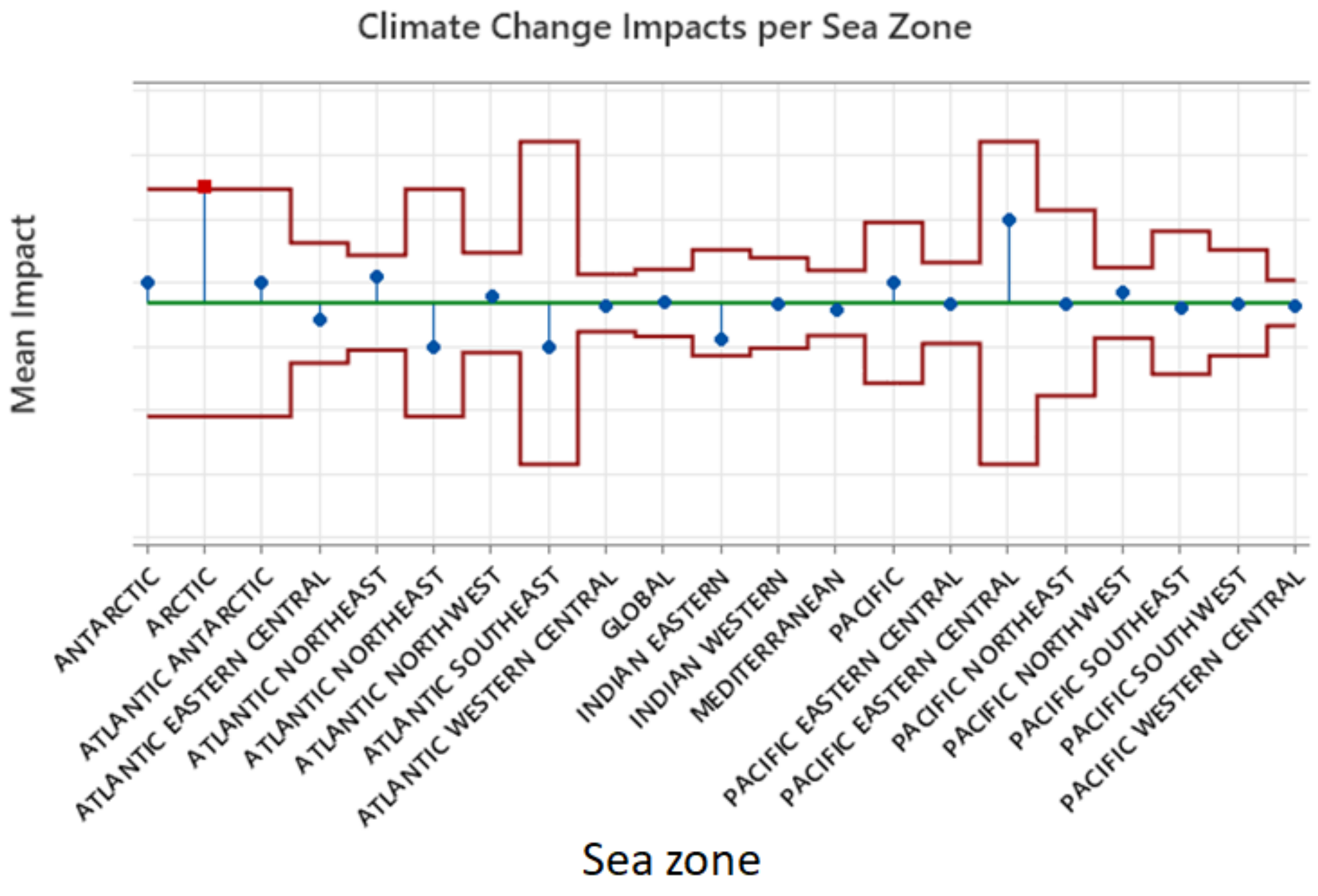

a

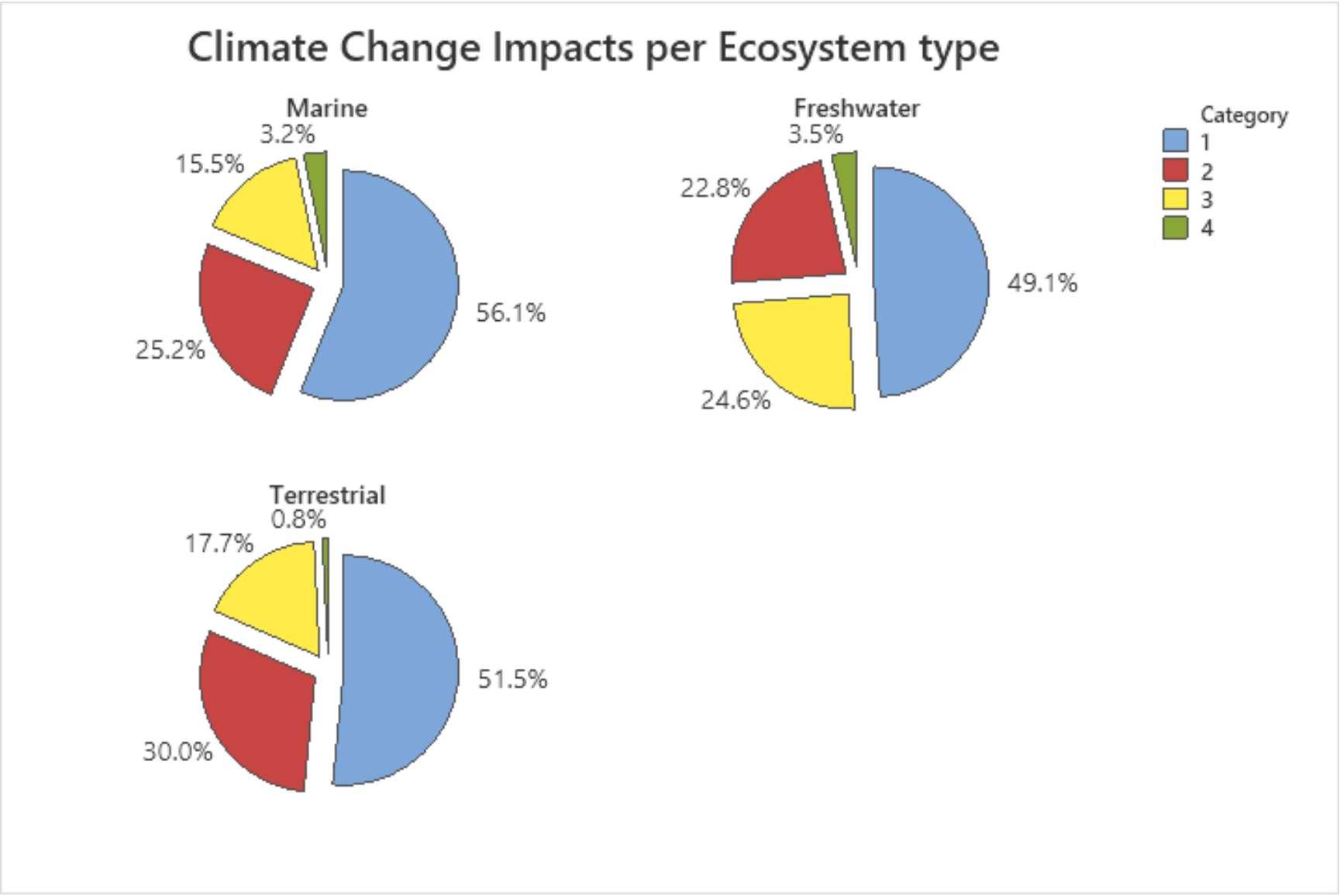

b

***Figure 2. a.*** *Analysis of mean impact per sea zone. Impacts are classified as 'negative' = 1, 'unclear' = 2, 'neutral' = 3, and 'positive' = 4. Analysis of means is a procedure to determine if individual factor level means are different from the grand mean (mean of all observations). The horizontal straight green line depicts the grand mean and red lines indicate the upper and lower 95% confidence interval of the effect on the grand mean per sea zone. Lines exceeding the upper or lower confidence interval indicate a significant effect.* ***b.*** *Pie chart of impact per ecosystem and their frequencies in the literature assessed.*

*3.1.3. Ecosystem Services*

In nearly all the ecosystem services investigated, climate change was found to have predominantly negative impacts. However, for 'genetic and biochemical resources' and 'cultural and spiritual significance' services, the majority of studies indicated unclear or inconclusive impacts (Fig. 3a). Among the ecosystem services examined, regulating services (especially climate and atmospheric regulation, and biological control and pollination) and soil-related services (i.e., soil formation and erosion control) experienced the most negative impacts (Fig. 3a). The highest frequency of neutral impacts was reported for biotic and abiotic material provision, freshwater provision and regulation, and food and nutrition provision (Fig. 3a). Very few studies report positive climate change impacts on an ecosystem service (Fig. 3a).

In studies that included other drivers, impacts of other drivers on ecosystem services were predominantly negative, except genetic and biochemical resources and soil-related services where the most frequent outcome was unclear (Fig. 3b). The waste and toxin mediation service category was reported with the same frequency to have negative as well as unclear impact (Fig 3b).

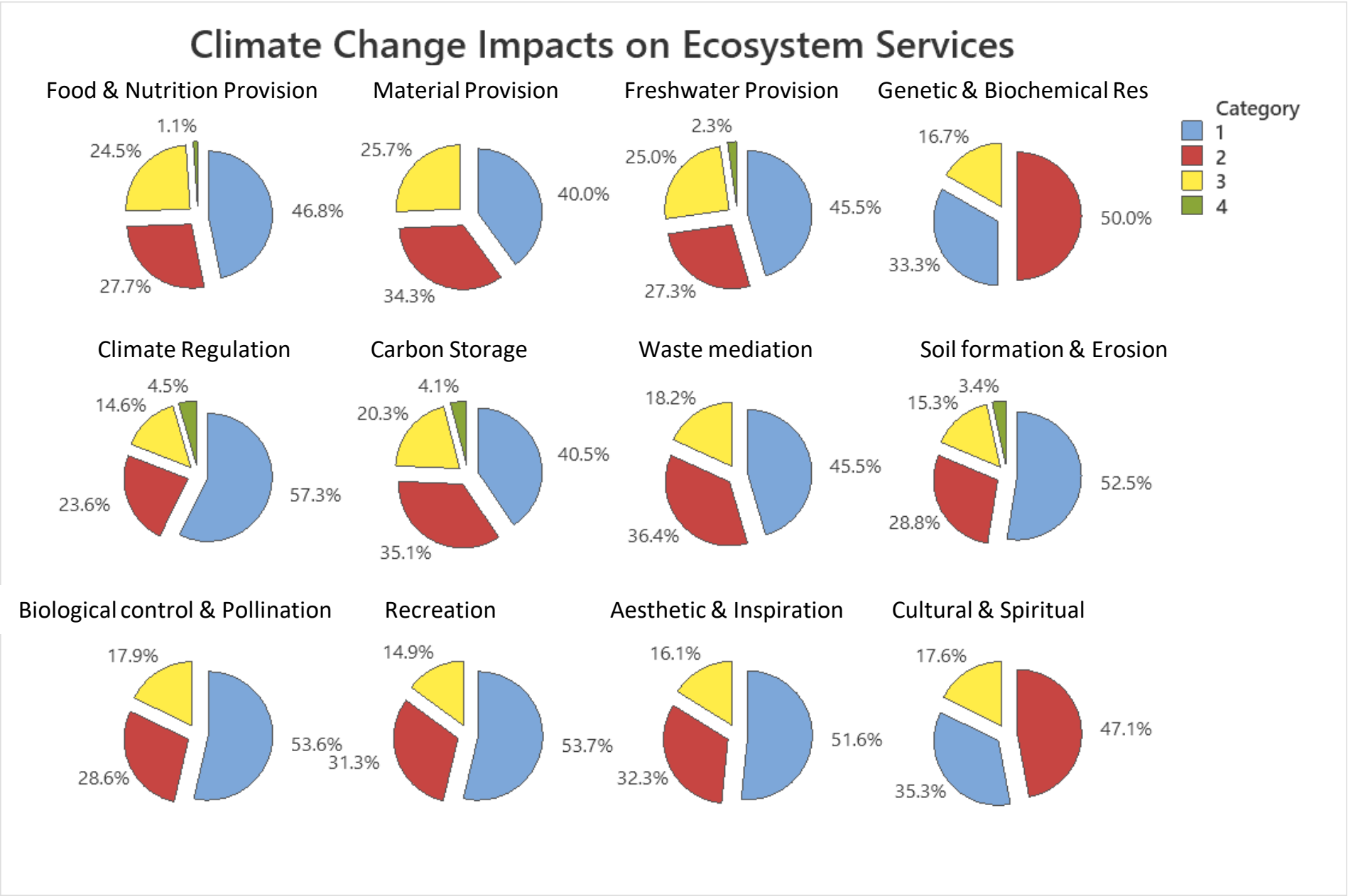

a

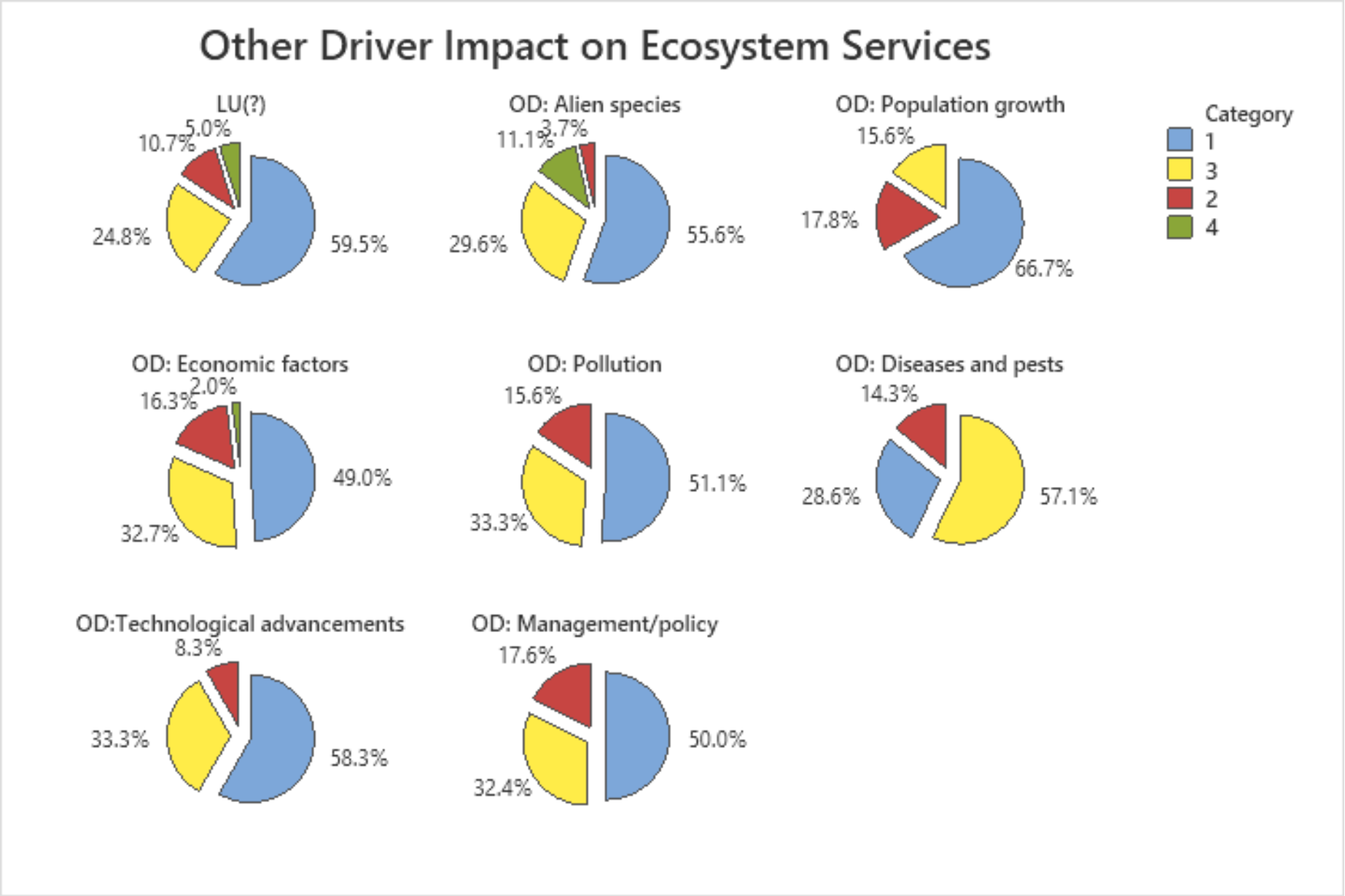

b

***Figure 3. a.*** *Pie chart of climate change impact per ecosystem service and their frequencies in the literature assessed. Impacts are classified as 'negative' = 1, 'unclear' = 2, 'neutral' = 3, and 'positive' = 4.*

***b.*** *Pie chart of other non-climatic driver impact per ecosystem service and their frequencies in the literature assessed.*

*3.1.4. Climate variables*

The climatic variable with the most frequent negative impact was ocean acidification, followed by sea level rise but generally negative impact frequencies are not very different between variables (Fig. 4a). Unclear and neutral impact frequencies are also similar across climate variables. Very few positive impacts are reported (Fig. 4a).

*3.1.5. Land use change*

In studies that included land use changes, impacts of land use changes were predominantly negative on ecosystem services except habitat protection that the impact was neutral (Fig. 4b). The most frequently negative impacts were recorded for deforestation, mining, urban expansion, coastal degradation, and wetland modification (Fig. 4b). Negative impacts on ecosystem services are recorded also for rewilding and habitat protection (Fig. 4b). In general land use changes exhibit considerable frequencies of neutral effects on ecosystem services (Fig. 4b).

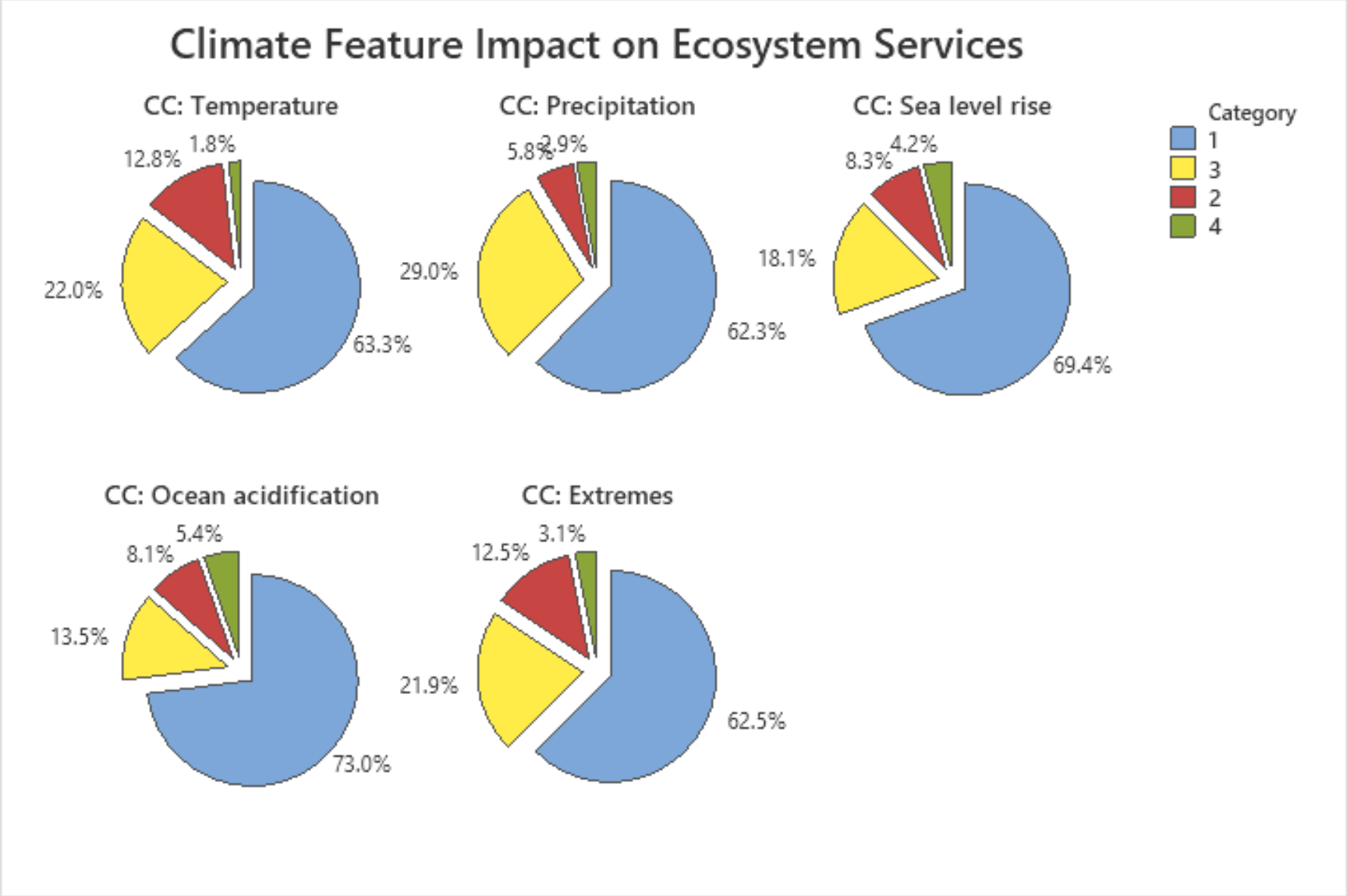

a

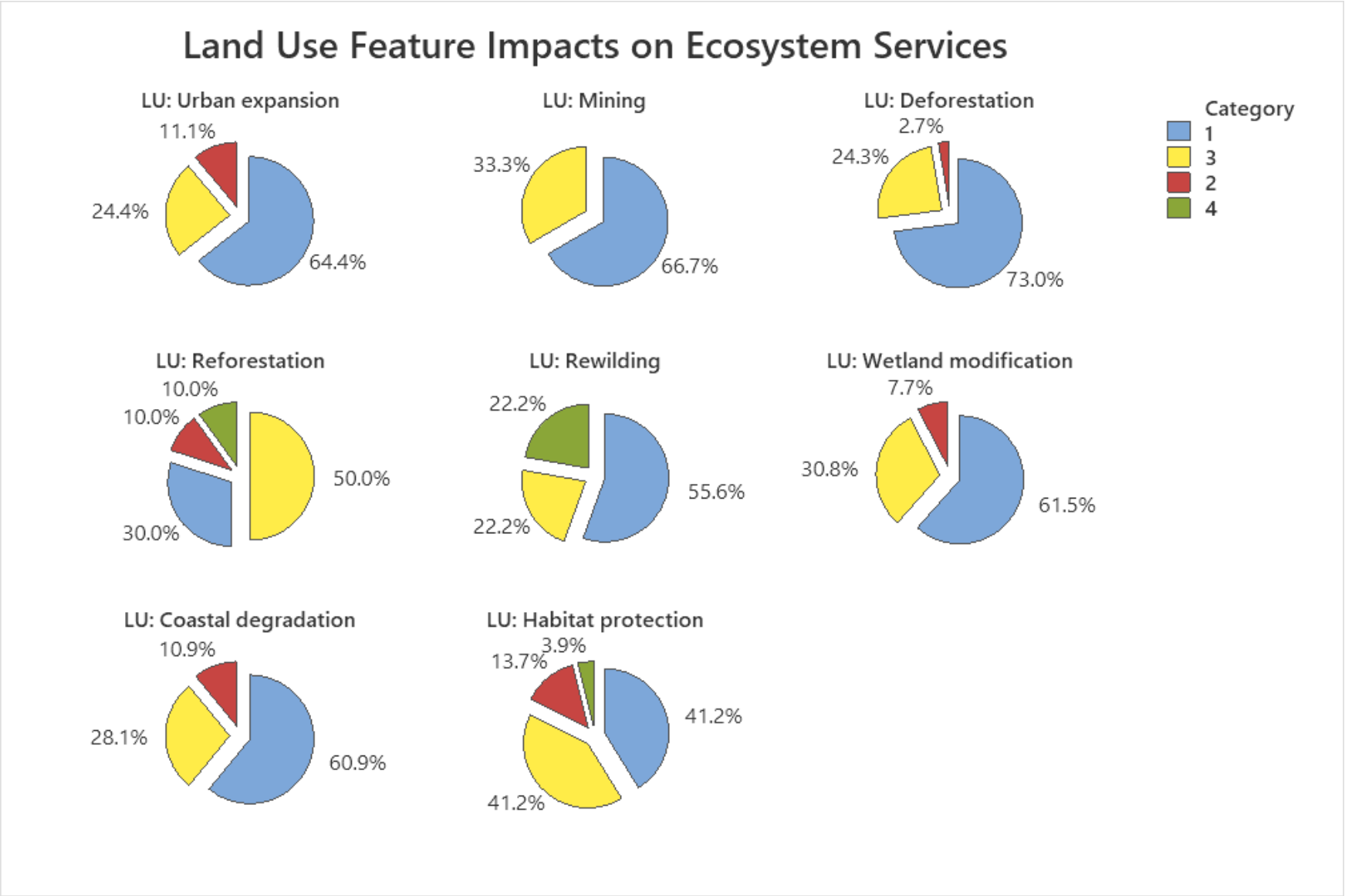

b

***Figure 4. a.*** *Pie chart of climate feature (variable) studied in the literature and the impact on ecosystem services and their frequencies in the literature assessed. Impacts are classified as 'negative' = 1, 'unclear' = 2, 'neutral' = 3, and 'positive' = 4.* ***b.*** *Pie chart of land use feature (variable) studied in the literature and the impact on ecosystem services and their frequencies in the literature assessed.*

### *3.2. Environmental analysis assessment*

Outputs of machine learning analysis combining all sea zone, ecosystem type, climate, land use, other driver variables indicated that negative ecosystem services impacts are primarily associated with coastal degradation, sea level rise, habitat protection and wetland modification (Fig. 5a). Population growth, policy or management interventions, and pollution followed (Fig. 5a). Most variables with high classification accuracy for negative climate change impacts on ecosystem services are variables related with land use changes followed by other driver variables with sea level rise as the only climate variable (Fig. 5a). No specific sea zones, or specific ecosystem, or ecosystem service type are found to be high accuracy predictors of negative impacts (Fig. 5a). Classification accuracy was close to 61.4% (varying between 60.08% - 62.1% based on the training to testing data partition) for the training and 57.5% (56.4% - 57.9%) for the test dataset respectively (Table 2a). Unclear impacts were primarily associated with precipitation (Fig. 5b). Coastal degradation, habitat protection, sea level rise, and population growth followed (Fig. 5b). Temperature and performing a global assessment had each a classification accuracy close to 20% each in relation to the variable with the highest accuracy i.e. precipitation. Policy interventions, invasive species, rewilding, and pollution were also found to be variables with relatively high classification accuracy (Fig. 5b). No specific sea zones, or specific ecosystem type, or ecosystem service are found to be high accuracy predictors of unclear impacts (Fig. 5b). Classification accuracy was 81% (80.04% - 81.7%) for the training and 74% (73.2% - 74.9%) for the test dataset (Table 2b). Neutral impacts were also primarily associated with precipitation (Fig. 5c). Food and nutrition provision, freshwater provision and regulation , population growth, physical and experiential recreation (e.g., tourism) , and economic factors were topping the list of high classification variables (Fig. 5c). Temperature, extreme weather phenomena, and Pacific North West sea zone were also recorded as variables with relatively high classification accuracy of neutral impacts. Aesthetic and inspirational value, biotic and abiotic material provision, and pollution were also included in the list (Fig. 5c). Classification accuracy was 51.6% (51.1% - 52%) in the training and 57.5% (57.2%-57.9%) in the test dataset (Table 2c).

There were insufficient data to fit a machine learning model regarding positive climate change impacts and model did not converge. Cases of positive impacts are qualitatively mentioned in the discussion.

*3.3. Methodological analysis assessment*

Outputs of machine learning analysis combining all methodological assessment methods and variables indicated that quantifying impact accuracy is primarily associated with the synergistic or antagonistic climate change – other driver (including land use change) interaction as well as the number of different climate features included in the study (Fig. 5d). The number of land use change types and the number of other driver variables, as well as the number of different ecosystem types studied were also important (Fig. 5d). Potential cumulative assessment of the other driver impact was also an important factor in classification accuracy. Interestingly the number of different ecosystem services investigated (denoted as ES richness) did not rank highly as a classification accuracy variable for climate change impacts on ecosystem services (Fig. 5d). Potential consideration of uncertainty or whether policy interventions were included in the analysis had also some role in the classification accuracy (Fig. 5d). Classification accuracy was 68.2% (67.6%-68.9%) in the training and 53.6% in the test dataset (Table 2c).

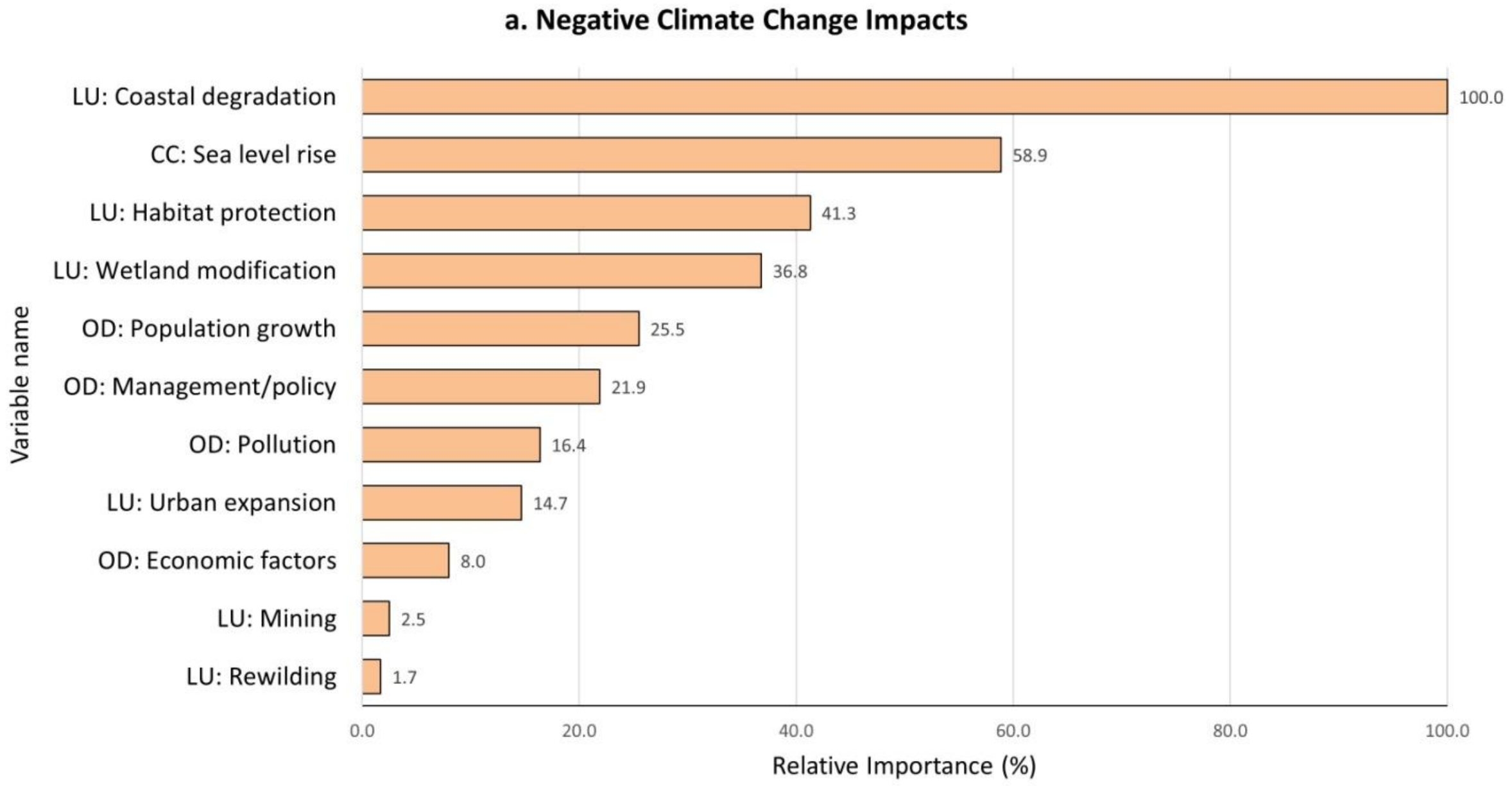

**b. Unclear Climate Change Impacts**

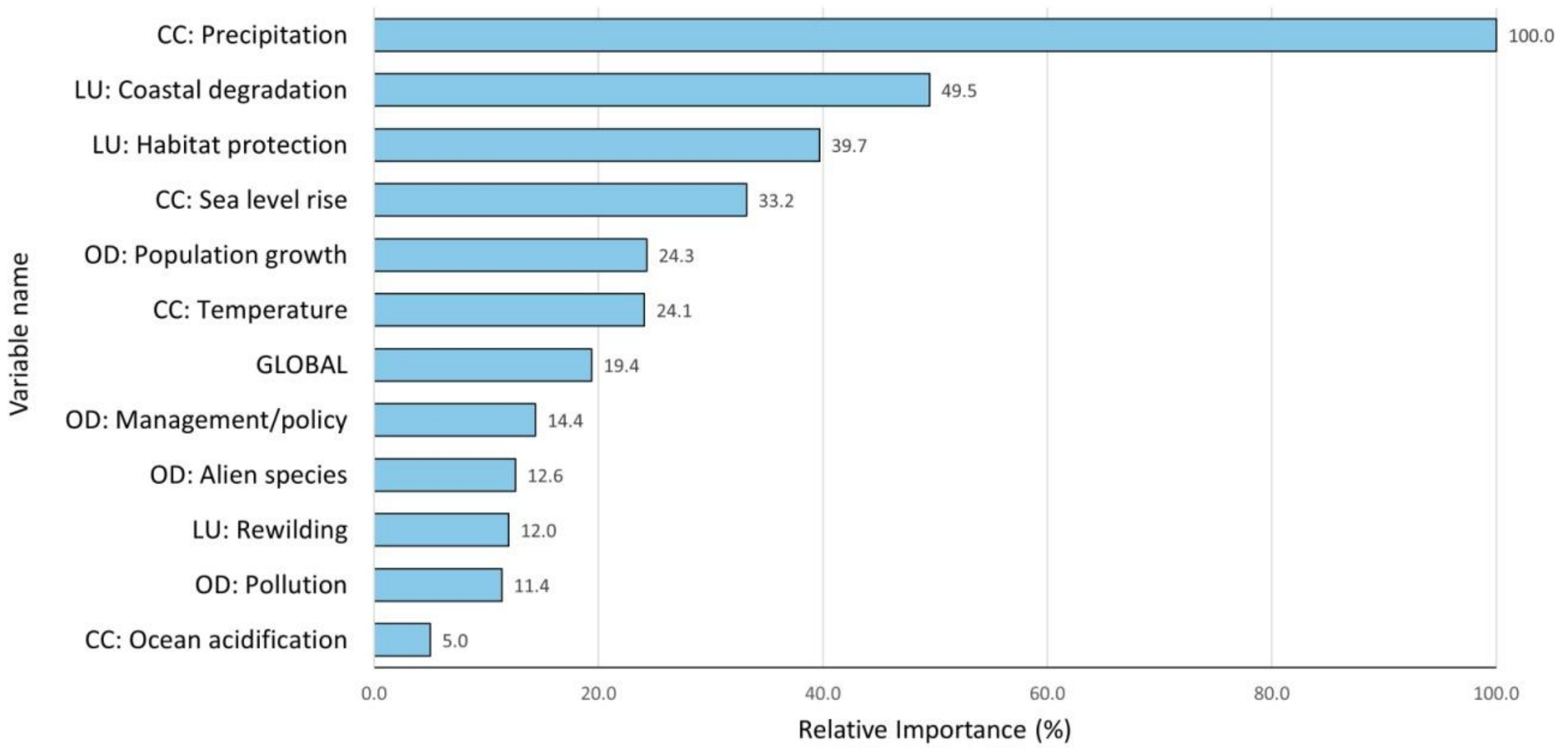

**c. Neutral Climate Change Impacts**

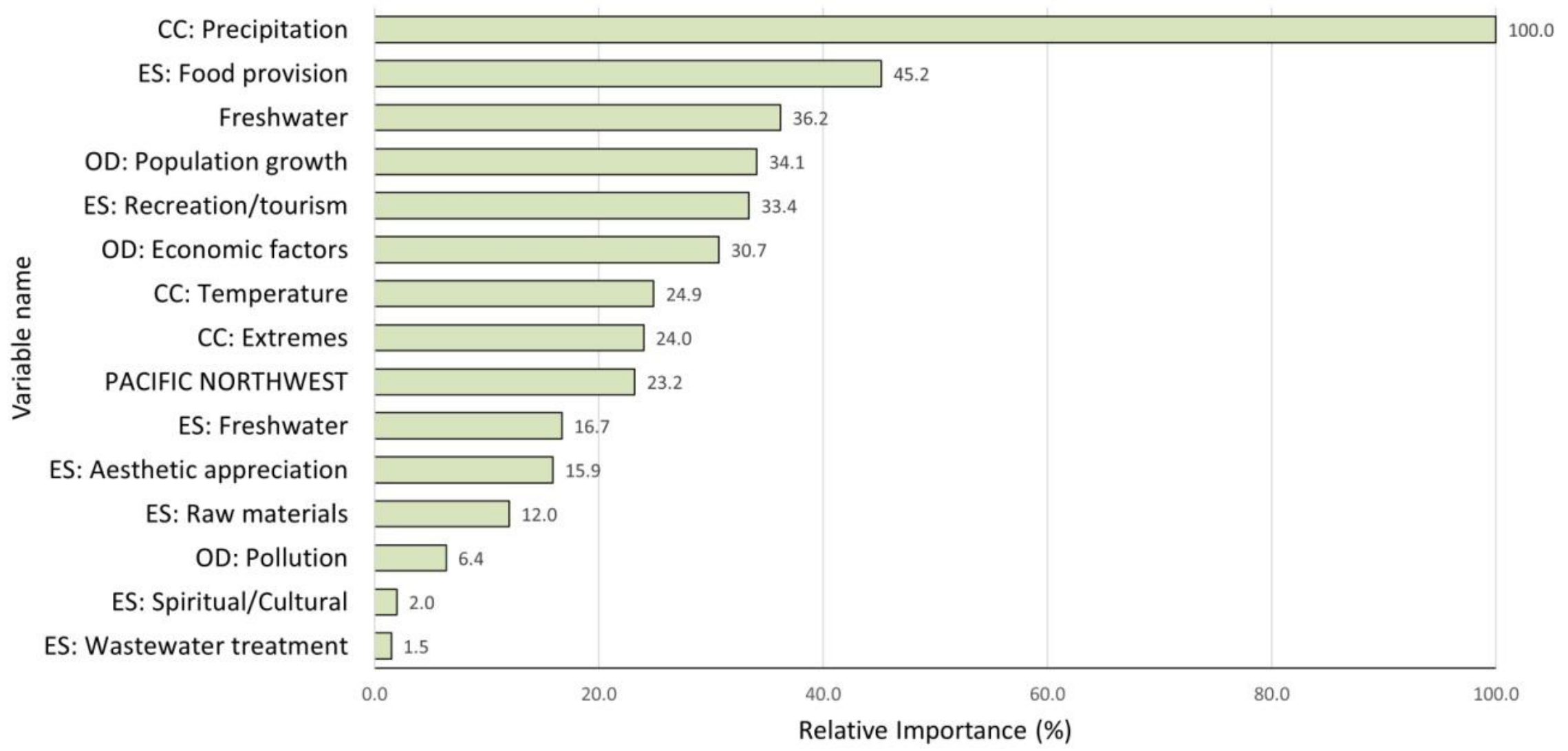

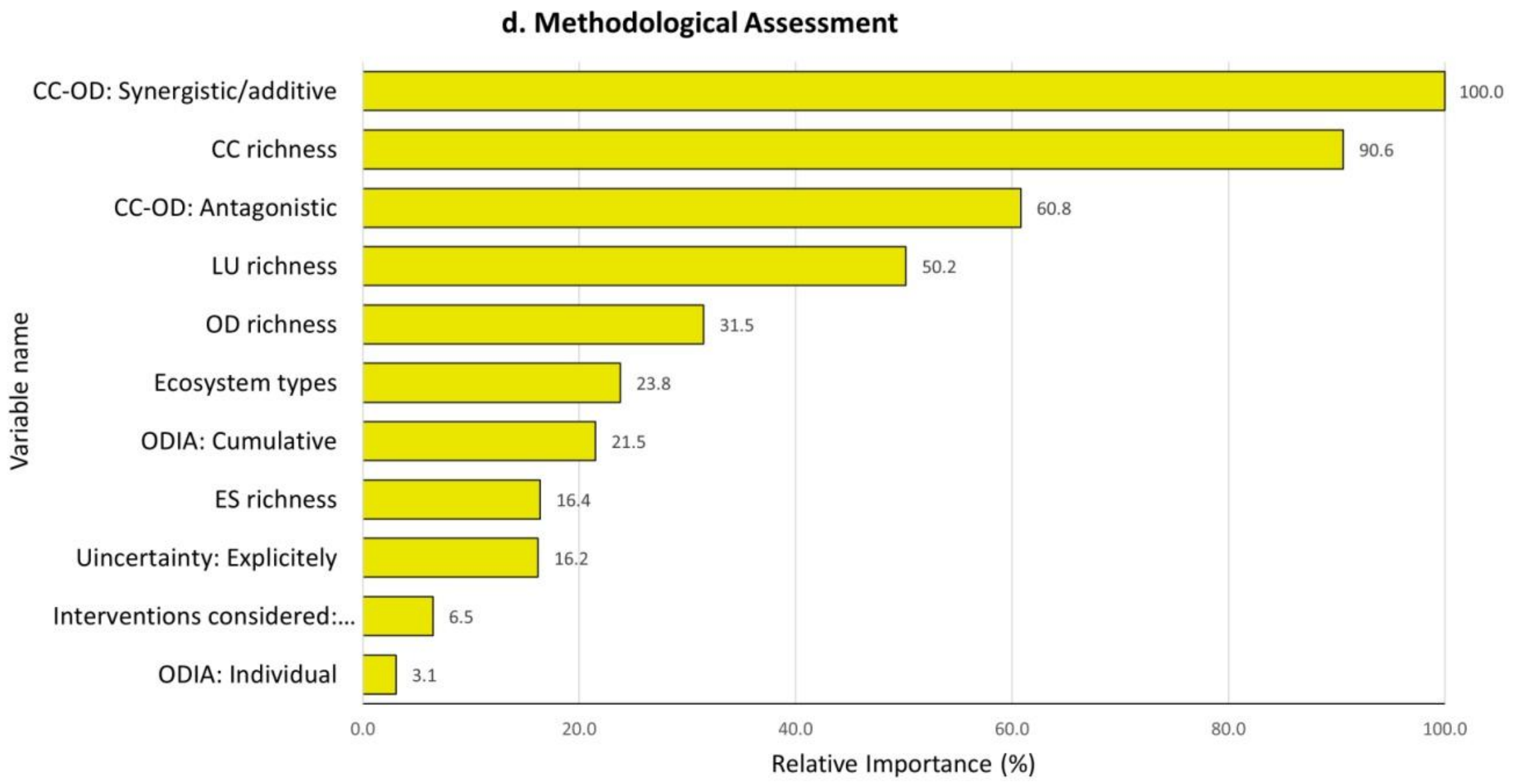

***Figure 5.*** *Machine learning analysis for understanding the impacts of climate change (CC) land use (LU) and other non-climatic drivers (OD) on island ecosystem services. Each impact outcome is analyzed separately to quantify the optimal classification accuracy variables explaining this outcome.* ***(a)*** *Negative impacts.* ***(b)*** *Unclear impacts.* ***(c)*** *Neutral impacts.* ***(d)*** *Methodological assessment of variables to quantify the classification accuracy of approaches and variables in detecting climate change impacts, regardless of what the outcome of the impact was.*

a. Negative impacts. Confusion matrix

| | Predicted Class (Training) | | | | Predicted Class (Test) | | | |
|---|---|---|---|---|---|---|---|---|
| **Actual Class** | **Count** | **1** | **0** | **% Correct** | **Count** | **1** | **0** | **% Correct** |
| 1 (Event) | 78 | 33 | 45 | 42.3 | 40 | 13 | 27 | 32.5 |
| 0 | 75 | 14 | 61 | 81.3 | 33 | 4 | 29 | 87.9 |
| All | 153 | 47 | 106 | 61.4 | 73 | 17 | 56 | 57.5 |

| **Statistics** | **Training (%)** | **Test (%)** |
|---|---|---|
| True positive rate (sensitivity or power) | 42.3 | 32.5 |
| False positive rate (type I error) | 18.7 | 12.1 |
| False negative rate (type II error) | 57.7 | 67.5 |
| True negative rate (specificity) | 81.3 | 87.9 |

b. Unclear impacts. Confusion matrix

| | Predicted Class (Training) | | | | Predicted Class (Test) | | | |
|---|---|---|---|---|---|---|---|---|
| **Actual Class** | **Count** | **1** | **0** | **% Correct** | **Count** | **1** | **0** | **% Correct** |
| 1 (Event) | 46 | 35 | 11 | 76.1 | 20 | 16 | 4 | 80.0 |
| 0 | 107 | 18 | 89 | 83.2 | 53 | 15 | 38 | 71.7 |
| All | 153 | 53 | 100 | 81.0 | 73 | 31 | 42 | 74.0 |

| **Statistics** | **Training (%)** | **Test (%)** |
|---|---|---|
| True positive rate (sensitivity or power) | 76.1 | 80.0 |
| False positive rate (type I error) | 16.8 | 28.3 |
| False negative rate (type II error) | 23.9 | 20.0 |
| True negative rate (specificity) | 83.2 | 71.7 |

c. Neutral impacts. Confusion matrix

| | Predicted Class (Training) | | | | Predicted Class (Test) | | | |
|---|---|---|---|---|---|---|---|---|
| **Actual Class** | **Count** | **1** | **0** | **% Correct** | **Count** | **1** | **0** | **% Correct** |
| 1 (Event) | 27 | 24 | 3 | 88.9 | 10 | 10 | 0 | 100.0 |
| 0 | 126 | 71 | 55 | 43.7 | 63 | 31 | 32 | 50.8 |
| All | 153 | 95 | 58 | 51.6 | 73 | 41 | 32 | 57.5 |

| **Statistics** | **Training (%)** | **Test (%)** |
|---|---|---|
| True positive rate (sensitivity or power) | 88.9 | 100.0 |
| False positive rate (type I error) | 56.3 | 49.2 |
| False negative rate (type II error) | 11.1 | 0.0 |
| True negative rate (specificity) | 43.7 | 50.8 |

d. Methodological assessment. Confusion matrix

| | | Predicted | | | | |
|---|---|---|---|---|---|---|
| **Sample** | **Observed** | **1** | **2** | **3** | **4** | **Percent Correct** |
| Training | 1 | 76 | 8 | 0 | 0 | 90.50% |
| | 2 | 14 | 29 | 0 | 0 | 67.40% |
| | 3 | 20 | 6 | 0 | 0 | 0.00% |
| | 4 | 3 | 0 | 0 | 0 | 0.00% |
| | Overall Percent | 72.40% | 27.60% | 0.00% | 0.00% | 67.30% |
| Testing | 1 | 29 | 5 | 0 | 0 | 85.30% |
| | 2 | 4 | 19 | 0 | 0 | 82.60% |
| | 3 | 9 | 1 | 0 | 0 | 0.00% |
| | 4 | 3 | 0 | 0 | 0 | 0.00% |
| | Overall Percent | 64.30% | 35.70% | 0.00% | 0.00% | 68.60% |
| Statistics | Training | Cross Entropy Error | 129.81 | Testing | Cross Entropy Error | 59.158 |
| | | Incorrect Predictions % | 32.70% | | Incorrect Predictions % | 31.40% |
| **AUC** | **CCI(-)** | **CCI(?)** | **CCI(0)** | **CCI(+)** | | |
| | 0.719 | 0.886 | 0.674 | 0.839 | | |

***Table 2.*** *Classification accuracy of machine learning analysis.* ***a.*** *Negative impacts* ***b.*** *Unclear impacts* ***c.*** *Neutral impacts* ***d.*** *Classification accuracy of methodological variables in detecting impact. AUC refers to the Area Under the Curve metric of model assessment.*

## 4. Discussion

Findings from the quantitative review conducted here indicate that land use changes and other drivers' variables are the highest accuracy variables, from the ones examined, for quantifying climate change impacts on island ecosystem services. In addition, climate change impacts are better quantified when including also land use variables (Sejr et al., 2024). Land use changes are more important for the assessment of climate change impacts on island ecosystems than changes in climate variables *per se*. In addition, this is applicable worldwide in islands, as there is no geographic location in terms of sea zone with more pronounced impacts. In some islands, as human population increases, demand for agricultural land is a major driver of natural habitat loss (Lee et al., 2015; Mayorga et al., 2022). In tourist destinations, human activities in islands are usually close to the coast and coastal pressures and degradation is also conducted for tourism (Merven et al., 2023; Morrissey et al., 2023; Senevirathna et al., 2018). Natural land conversion for artificial or agricultural land use is causing species loss and disrupting ecosystem functioning (Fois et al., 2018), therefore affecting the flow of ecosystem services.

### *4.1. Sea zone, Ecosystem Type, and Ecosystem Services*

Determining whether marine ecosystem types are more impacted than terrestrial ones within the same island or study area is generally challenging (Antão et al., 2020). However, the type of ecosystem *per se* was a relatively low predictor of the impact indicating that climate change impacts are not more pronounced in a specific ecosystem over another. Similarly, analysis of impacts per sea zone and machine learning variable importance outputs indicate that there is no strong evidence that the impacts of climate change or other drivers are more pronounced in a particular sea zone. Thus there is no evidence that a specific location is undergoing a higher impact burden. This does not imply that the form of climate change is identical across regions; form example some regions may receive higher or lower precipitation or more extremes (Bhatti et al., 2021; Homar et al., 2010; McGree et al., 2019). No particular ecosystem service category is under a severe negative climate change impact when all environmental factors are accounted for as indicated by machine learning variable importance outputs. This is in spite of the negative impacts recorded as the most frequent outcome in the studies investigated for most ecosystem service categories. Specific ecosystem services, such as food and nutrition provision and physical and experiential recreation, are associated with neutral impacts in studies that were included. A study on the perception of climate change impacts on ecosystem services in very different areas, from the island of Socotra to Senegal, underlines the importance of socio-ecological temporal dynamics to better understand how these vary geographically and according to prevailing activities (Mehring et al., 2017).

### *4.2. Climate features*

Sea level rise was the climatic variable with the most pronounced effect on negative climate change impacts on ecosystem services. Sea level rise can significantly impact low-lying coastal and intertidal habitats, leading to widespread flooding and accelerated coastal erosion. Small islands, reefs, and atolls are biodiversity hotspots or home to rare endemic species that are particularly vulnerable to habitat loss. Climate change acting through sea level rise results in habitat loss further exacerbating the land use change – climate change interaction (Roy et al., 2023). Inclusion of precipitation in a study was associated with unclear or neutral impacts. This, in part, can be explained by higher or lower precipitation volume, or more frequent precipitation extremes recorded in different islands (Bhatti et al., 2021; Homar et al., 2010; McGree et al., 2019), thus a non-consistent trend may result in a non-consistent impact outcome.

### 4.3. *Land Use features & Other Drivers*

Land use characteristics are dominating the list of negative climate change impacts and include the negative impacts of coastal degradation, wetland modification, urban expansion and mining. They also include maintaining or increasing natural land use types such as habitat protection and rewilding. The overall impact was negative even in cases that habitat protection and rewilding was studied indicating that their effects were moderated. This needs to be seen in all cases synergistically with the climatic change impact (Brodie 2016; Fox et al., 2014). It is noteworthy that policy or management changes were ranked highly as an explanatory factor of climate change impacts. Policy and management (Butler et al., 2014; Jupiter et al., 2014) were only somewhat important when impacts were unclear, while had low relative importance in inverting negative or maintaining neutral impacts. The efficacy of policy or management actions deployed to mitigate those impacts is rarely quantified, synthesized and thus poorly understood (Milanese et al., 2011). Potential need for adaptation of policies to islands, as well as their efficacy *on* islands, remains insufficiently clear. In the European Union, for example, the need for adapting Common Agricultural Policy, bringing back nature to all ecosystems (European Commission, 2022), and the Biodiversity Strategy 2030 (European Commission, 2021), is lacking in the islands context. At more local scales, the effects of nature-based solutions (Châles et al., 2023) are also hard to assess as these are rarely accounted for in the literature, despite the positive effect found by the few available island-specific studies (Antonsen et al., 2022; Kiddle et al., 2021; Zari et al., 2019). Stakeholder engagement was found to be critical in enhancing the implementation of nature-based solutions in islands (Chen et al., 2024; Grace et al., 2021).

### *4.4. Positive impacts*

There are few cases documenting positive impacts of climate change. Yet, these results should be interpreted cautiously as the sample size of positive climate change impacts is small. Such examples include the northward expansion of *Conocarpus erectus* (also known as button mangrove), a species with high potential for carbon sequestration (Ochoa-Gómez et al., 2021). Similarly, *Avicennia germinans (L.)*, the most cold-tolerant mangroves in North America, are expected to increase in productivity given continued warming, and infrequent frosts (Steinmuller et al., 2022). *Posidonia oceanica* (L.) Delile seagrass meadows are highly productive coastal marine ecosystems that also provide multiple ecosystem services. Their abundance in the Balearic island of Cabrera is favoured by elevated solar irradiance and increased air temperature (Leiva-Dueñas et al., 2020). Similarly, coastal freshwater ecosystems in southern Victoria Island may become more productive and global sea level rise may thus trigger a new pathway to an increasingly greener Arctic (Kivilä et al., 2022).

### *4.5. Methodological assessment*

The diversity of climate variables is found as an important predictor of the climate change impact outcome. Similarly, more land use as well as other driver features results in much higher accuracy of identifying impacts. This naturally leads to an increased complexity in the analysis, with more simple studies being more likely to miss some actual impacts (Evans et al., 2013). However, there are subtleties to consider when including non-climatic drivers, such as the way they interact with each other and with climate (Redlich et al., 2022). Understanding whether they act synergistically or antagonistically, additively, or cumulatively, is crucial for effective awareness, adaptation and mitigation strategies (Mayorga et al., 2022). However, these interactions are not consistently or clearly defined. For example, in the small islands of the Western Indian Ocean, these limitations reverberate in ineffective communication with local populations (Newman et al., 2020) or, in Hokkaido, in an unsuitable fishery management (Kaeriyama et al., 2012).

Handling uncertainties did not rank highly as a methodological variable for impact classification. In several cases uncertainty is unaccounted for and thus the potential effect in the result derives mostly from relatively few cases where uncertainty assessment was explicitly accounted for (Katsanevakis and Moustakas 2018; Runting et al., 2017). In most cases uncertainty is either ignored or simply mentioned. A common cause of uncertainty of island ecosystem services is the limited spatiotemporal resolution of climate or ecosystem services data (Sieber et al., 2021; Vogiatzakis et al., 2023). This is especially true for the smallest islands, where most assessments rely on coarse-resolution data that may not capture localized variations, which are critical for island-specific analyses (Fain et al., 2018; Vogiatzakis et al., 2023). Thus, there is a need for downscaling data to the island level (Chou et al., 2020; Daliakopoulos et al., 2017; León et al., 2021).

*4.6. Limitations*

Impact evaluation as conducted here is subjective to the expert evaluator (reader). Despite cross-validating impact evaluations between readers, the approach is limited by the subjective expert evaluation. Scale issues are hard to address as studies may include part of the land or sea area of an island while others may include the whole island or islands/study areas of profoundly different surface area (Allen, 2017; Calado, Bragagnolo, Silva, & Pereira, 2014). Impacts can vary with scale and a negative impact at a local scale may be a positive one at a coarser scale and *vice versa* . Determining the optimal interaction scale (Moustakas, Daliakopoulos, & Benton, 2019) between climate and ecosystem services may also vary across ecosystem service types. All these scale-related issues are unaccounted for in this study. Publication year may potentially play a role (Kadykalo et al., 2019) as the term 'ecosystem services' is appearing more often in publications in recent years thereby potentially increasing inclusion of the term in recent publications that ecosystem services are not the core topic and reciprocally not appearing in older publications that the terms was not as widely used. However publication year does not necessarily correspond to the time of study when the impact occurred (Moustakas & Karakassis, 2005). There are also studies spanning for several years. Thus impacts recorded here are not time-adjusted.

*4.7 Management Implications*

Our findings have implications for climate change impact assessment, climate change adaptation and climate change mitigation. In islands the key to modulating climate impacts are land use changes. This is potentially due to more restricted surface area in comparison to mainland (Chen, 2025; Vogiatzakis et al., 2023). Thus tailored actions to island-specific pressures are needed. As pressures differ among islands, management needs to be locally adjusted and customised, with detailed assessments of coastal development, population growth, and other island-specific factors and pressures. This implies that island-specific management strategies and climate-resilient policy measures for island ecosystems are needed that may differ from regional, national or EU-level policies (Albrecht, Belinskij, & Heikkilä, 2025). In addition, policymakers need to monitor and evaluate the outcomes of conservation policies (e.g., EU biodiversity strategy) on islands, ensuring continuous adaptation and stakeholder involvement to improve efficacy (Ruiz et al., 2023). Quantifying land use changes that amplify climate impacts as well as land use changes that mitigate climate impacts should be a priority. Legislation, management, and motives to civil stakeholders should be co-created so as to minimize land use changes activities that maximize impacts of the most negative drivers quantified such as coastal degradation, wetland modification and human population growth in islands. Reciprocally habitat protection and reforestation can minimize climate impacts with an additional potential for more sustainable tourism and should be further promoted (Mason & Neumann, 2024). Research and managers need to assess land-use change alongside climate drivers (e.g., sea-level rise) to fully capture impact pathways and design more integrated adaptation and mitigation strategies. Management strategies should incorporate uncertainty analyses (e.g., scenario planning, robust monitoring) to better cope with data gaps, variable island conditions, and nonlinear responses of ecosystem services (Walther et al., 2025).

## 5. Conclusions

Land-use features (e.g., coastal degradation, wetland modification) emerged as top predictors of climate change impacts on island ecosystem services*, often more influential than climate variables alone.* No particular sea zone or single ecosystem type stood out as *uniformly* more vulnerable; however, local land-use activities (e.g., *urban expansion, mining*) often intensified climate impacts.

Negative impacts on ecosystem services were significantly tied to coastal degradation, deforestation, and wetland modification. Conversely, habitat protection and rewilding sometimes mitigated negative effects but not always strongly enough to offset climate stressors. Studies that included multiple climate features (e.g., temperature, precipitation, ocean acidification, extremes) had higher accuracy in identifying negative impacts. Population growth, economic factors, and technological change ranked highly among “other driver” predictors of negative outcomes. While policy/management variables appeared as moderately important predictors, the effectiveness of such interventions (e.g., nature-based solutions, local management plans) was rarely quantified. Spatial resolution and data limitations in islands make it harder to detect changes in ecosystem services over time. Uncertainty handling was often minimal; many studies either ignored or only qualitatively mentioned uncertainties.Understanding climatic, land use, and other non-climatic drivers' interactions is crucial for effective awareness, adaptation, and mitigation strategies.

## Author Contributions

A.M., S.Z-S., M.T. and G.Z. conceptualised and designed the research; S.Z-S., M.T. and G.Z. collected the data; all co-authors contributed to reading papers and recording variables; A.M. and G.Z. analysed the data; A.M., G.Z., and C.Z. were the core writing team; S.Z. G.Z. and A.M. visualized the results; all co-authors commented on the manuscript.

## Data availability

The results of the literature review are publicly available at https://doi.org/10.5281/zenodo.10518892

## Acknowledgements

Comments of the handling editor, Paulo Pereira, and two anonymous reviewers greatly improved an earlier manuscript draft. This research was supported by the COST Action SMILES (CA21158): Enhancing Small-Medium IsLands resilience by securing the sustainability of Ecosystem Services.